# Quantum interference in heterogeneous superconducting-photonic circuits on a silicon chip

C. Schuck[1], X. Guo[1], L. R. Fan[1], X. S. Ma[1,2], M. Poot[1] and H. X. Tang[1*]

[1]*Department of Electrical Engineering, Yale University, New Haven, CT 06511, USA*
[2]*Institute for Quantum Optics and Quantum Information, Austrian Academy of Science, Vienna, Austria*
*Correspondence should be addressed to H.X.T. (email: hong.tang@yale.edu)*

Quantum information processing holds great promise for communicating and computing data efficiently. However, scaling current photonic implementation approaches to larger system size remains an outstanding challenge for realizing disruptive quantum technology. Two main ingredients of quantum information processors are quantum interference and single-photon detectors. Here we develop a hybrid superconducting-photonic circuit system to show how these elements can be combined in a scalable fashion on a silicon chip. We demonstrate the suitability of this approach for integrated quantum optics by interfering and detecting photon pairs directly on the chip with waveguide-coupled single-photon detectors. Using a directional coupler implemented with silicon nitride nanophotonic waveguides, we observe 97% interference visibility when measuring photon statistics with two monolithically integrated superconducting single photon detectors. The photonic circuit and detector fabrication processes are compatible with standard semiconductor thin-film technology, making it possible to implement more complex and larger scale quantum photonic circuits on silicon chips.

Proof-of-principle experiments have shown that quantum information processing has great potential for solving certain computational tasks, which are intractable with classical means [1]. Among the various approaches integrated quantum photonics has emerged as a particularly interesting one for realizing optical quantum simulations [2,3,4,5], quantum information processing [6,7,8,9] and communication [10,11]. However, scaling current quantum technology to larger system sizes remains a significant challenge due to the demanding requirements for high-fidelity signal processing at single-photon levels.

Advanced nanofabrication techniques have proven invaluable for ensuring scalability of electronic components used in classical information technology [12]. The corresponding complementary metal oxide semiconductor (CMOS) fabrication recipes have recently also been employed for realizing both nanophotonic waveguides [13] as well as superconducting single-photon detectors (SSPD) [14] on silicon chips. As most linear optics quantum logic schemes rely on non-classical interference and single-photon detection [15,16] it is crucial to realize both of these ingredients on a common scalable platform. Here we demonstrate such a quantum information processing platform by combining SSPDs with integrated silicon nitride photonic circuits to measure high-visibility quantum interference directly on-chip.

Highly efficient single-photon detection has previously been achieved with fiber-coupled SSPDs [17], which have found many exciting applications [18,19,20]. For integrated photonic technology, detection of photons inside a waveguide directly on-chip is required because it eliminates the chip-to-fiber interface, which is often a bottleneck in photonic device packaging. Optimal performance in this regard is achieved with nanowire SSPDs in traveling wave geometry [21,22]. This design is an excellent choice for integrated nanophotonic applications because large numbers of these compact detectors can be embedded directly in optical waveguide circuits [23]. On-chip detection efficiencies up to 90% [24,25] (system detection efficiencies up to 10% [26,27]) at visible as well as telecom wavelengths have been demonstrated with such waveguide coupled SSPDs. Furthermore, these detectors can operate at GHz rates, achieve <20 ps timing accuracy, sub-Hz dark count rate and extremely low noise equivalent powers down to the $10^{-20}$ W Hz$^{-1/2}$-level [24,28].

Quantum interference can be observed when two indistinguishable photons impinge simultaneously on the inputs of a 50:50 beam splitter, i.e. the probability of finding individual photons in separate output modes vanishes. This non-classical effect was first observed by Hong, Ou, and Mandel (HOM) [29] and is the consequence of destructive interference of the probability amplitudes corresponding to both photons being transmitted / reflected at the beam splitter [30]. A waveguide implementation of an optical beam splitter is an optimal choice for realizing spatial mode matching [31,32,33,34], which is one of the limiting factors for achieving high-visibility quantum interference in free-space optics experiments.

The integration of photonic circuits and detectors on a silicon chip for demonstrating quantum interference has previously been attempted with surface plasmon polariton devices. Two-plasmon quantum interference with 93% visibility on a beam splitter has been demonstrated with off-chip detectors and for photons at visible wavelengths [35], which are not compatible with existing optical communication networks. Notably, the integration of plasmonic directional couplers with superconducting detectors on the same chip proved challenging and reduced the interference contrast below the classical limit [36].

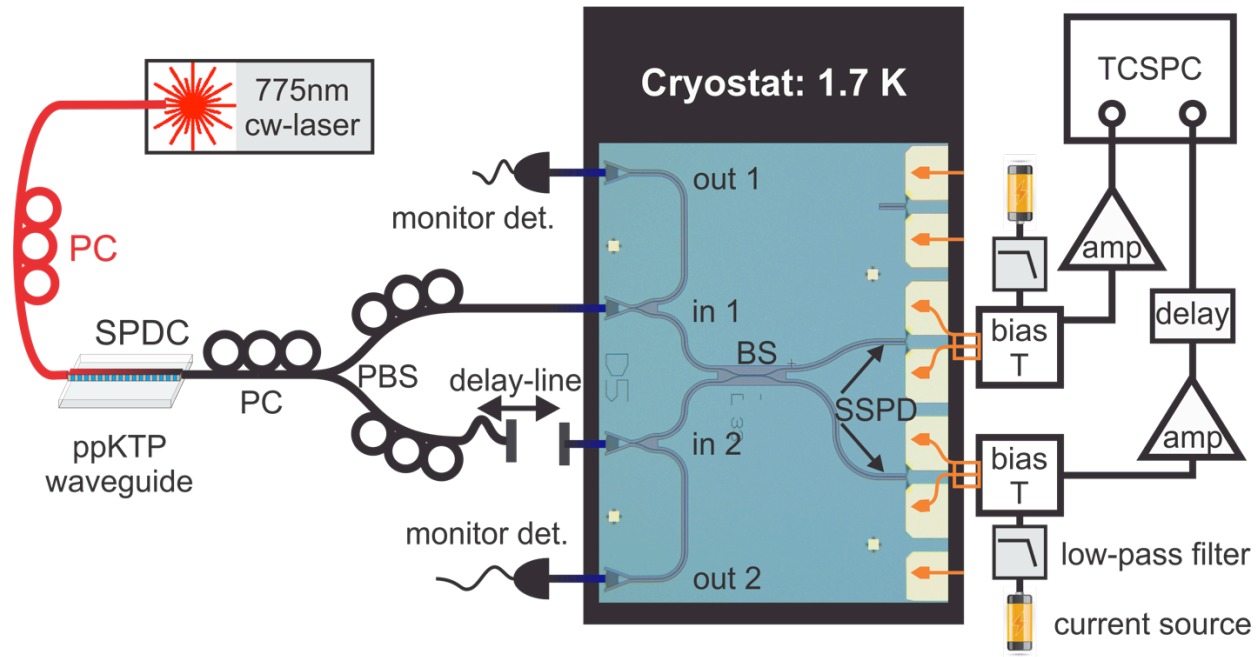

**Fig. 1. Schematic of the on-chip quantum interference measurement setup.** A 775nm continuous wave (cw) diode laser is used as a pump for generating orthogonally polarized 1550 nm photon pairs via type-II spontaneous parametric down in a 10.5 mm-long fiber-coupled periodically poled KTP waveguide. Input polarizations to the ppKTP source and a polarizing fiber beam splitter (PBS) are adjusted with fiber polarization controllers (PC) for optimal SPDC efficiency and deterministic splitting of photon pairs into separate PBS-output modes, respectively. Temporal delay between photons in separate output modes is set with a fiber-free-space-fiber optical delay line. Light is then guided into a closed-cycle cryostat where it is coupled from an optical fiber array into on-chip SiN photonic circuits via optical grating couplers at *in 1* and *in 2* (optical micrograph). The alignment of the chip to the fiber array with low-temperature nanopositioners (attocube) is aided by monitoring the optical transmission at the auxiliary ports *out 1* and *out 2*. Photons interfere at a 33 μm long directional coupler (beam splitter, BS) with 400 nm gap between waveguides in the coupling region. Superconducting single-photon detectors (SSPD) on top of the BS's output waveguides are supplied with 10-15 μA bias from a low noise current source and read out via a rf-probe connected to a bias-T outside the cryostat. After signal amplification (PPL 5828 & RF-Bay LNA-4050), photon statistics are recorded with a time-correlated single-photon counting unit (TCSPC, PicoHarp 300).

Here we integrate low-noise niobium titanium nitride (NbTiN) nanowire SSPDs with dielectric silicon nitride (SiN) photonic circuits on a silicon chip. Using photons from spontaneous parametric down conversion we measure quantum interference with 97% visibility directly on-chip. Our circuit-detector approach is fully compatible with scalable, high-yield semiconductor microfabrication processes.

## Results

**Experimental setup.** The experimental setup for measuring Hong-Ou-Mandel interference with SSPDs directly on-chip is shown in Fig. 1. We produce energy-time correlated photon-pairs via the process of type-II spontaneous parametric down conversion (SPDC) in a periodically poled potassium titanyl phosphate (ppKTP) crystal-waveguide [37]. A continuous wave 775 nm pump laser is coupled directly from an optical single mode fiber into the ppKTP-waveguide and the generated 1550 nm photon pairs are collected into a single mode fiber. In type-II SPDC the generated photons of one pair have orthogonal polarization. Hence we use a fiber polarization beam splitter (PBS) to deterministically separate photons of each pair. To do so, we optimize the efficiency of the source by using a 1550 nm telecom laser, which we send via both inputs/outputs of the fiber PBS through the ppKTP waveguide, i.e. in reverse, and monitor the second harmonic generation (SHG) of 775 nm light. First we maximize the SHG power to optimize spatial alignment of the crystal waveguide's input/output fibers and the crystal temperature for a given wavelength and ppKTP waveguide geometry. We then minimize the SHG power using only the polarization controller between the ppKTP crystal and the fiber PBS (see Fig. 1), thus ensuring orthogonal polarizations (H/V) at the PBS. Switching back to the 775 nm pump laser we first adjust the polarization controllers between 775 nm laser and ppKTP waveguide to maximize photon pair generation efficiency. Then we adjust the polarization controllers behind the fiber PBS such that we achieve optimal coupling to the TE-mode of the on-chip waveguide. We use a 1064 nm long-pass and a 1550 nm band-pass filter, which efficiently suppress 775 nm, pump light in combination with on-chip grating couplers, effectively acting as additional band-pass filters. We introduce an optical delay line in one of the output ports of the fiber PBS that allows us to scan the relative arrival time between two photons of a pair at the silicon chip.

The chip with approximately one hundred photonic integrated circuits (PIC) and twice as many detectors (SSPDs) is mounted inside a closed-cycle cryostat, which provides continuous cooling to 1.7 K with less than 10 mK temperature variations [38]. We use a radiofrequency (rf) probe to make electrical contact to electrode pads on the chip, which connect to the SSPDs. The rf-probe lines are wired to a bias-T for supplying current from a low noise source to the nanowires as well as reading out the voltage pulses upon photon detection by the SSPD [23]. Photons from the down conversion source are delivered to the on-chip PICs via an optical fiber array. Coupling loss from optical fibers to the on-chip waveguides is calibrated independently for each device input via monitor ports (see Fig. 1) and is usually around 10 dB. The two device layouts shown in Fig. 1 and Fig. 2a facilitate detector characterization, optical path-length measurements and fiber-to-device alignment. However, using 3 dB of the signal (idler) photons per input port for calibration purposes also leads to a quadratic decrease in the coincidence detection rates from correlated photon pairs. Future device designs could benefit from omitting these monitor ports. The chip is mounted on a stack of low-temperature compatible translation stages which allow us to position different devices under the fiber array and rf-probe-assembly for testing.

We amplify SSPD output signals before recording their arrival times with a time-correlated single photon counting (TCSPC) unit. In offline analysis we can thus identify detection events from correlated photon pairs by comparing arrival time lists for each of the two SSPDs [32].

**Integrating photonic circuits and SSPDs**. The device geometry for an on-chip HOM-interference measurement is shown in Fig. 2a. Fabrication starts from a commercial SiN on silicon dioxide ($SiO_2$) on silicon (Si) wafer onto which we sputter a thin film of NbTiN. Subsequently we define electrode pads, nanowire detectors and waveguides in standard electron beam lithography followed by lift-off and dry-etching chemistry, respectively (see Methods). We use optical grating couplers to transmit light from the optical fiber array into waveguides of 1 μm width, designed for transverse electric (TE) single-mode propagation on-chip [24,25]. Photons are then guided to a beam splitter and detected by NbTiN nanowire SSPDs at the beam splitter's output.

The beam splitter is implemented as a directional coupler where two input waveguides are brought into close proximity over a coupling length $L_c$ (Fig. 2b). With the waveguides acting as polarizers and spatial mode filters, the photons are indistinguishable when they arrive at the directional coupler. Finite element (FEM) simulations of the coupling region show evanescent coupling between TE modes of 1 μm wide SiN-waveguides. The resulting symmetric (Fig. 2c) and anti-symmetric (Fig. 2d) hybrid modes have a refractive index difference of $\Delta n$ for a given gap. In simulations we find a coupling length of $L_c = \lambda/(4 \cdot \Delta n) = 28$ μm for a 400 nm gap to realize 50:50 splitting between the 330 nm high waveguides (see Fig. 2e). In calibration devices (Mach-Zehnder interferometers and beam splitters, see supplementary notes 1) we observe that a slightly larger coupler length of $L_c = 33$ μm is required to achieve a 50:50 splitting ratio, which accounts for waveguide asymmetry, width and gap-offsets, as well as additional coupling in the input and output region of the directional coupler, all of which are not taken into account in the FEM-simulations. Due to the high refractive index contrast of SiN on insulator ($SiO_2$), the footprint of our devices is approximately two orders of magnitude smaller than glass-based waveguide implementations and could in principle be even more compact if a smaller gap is chosen (see supplementary notes 1).

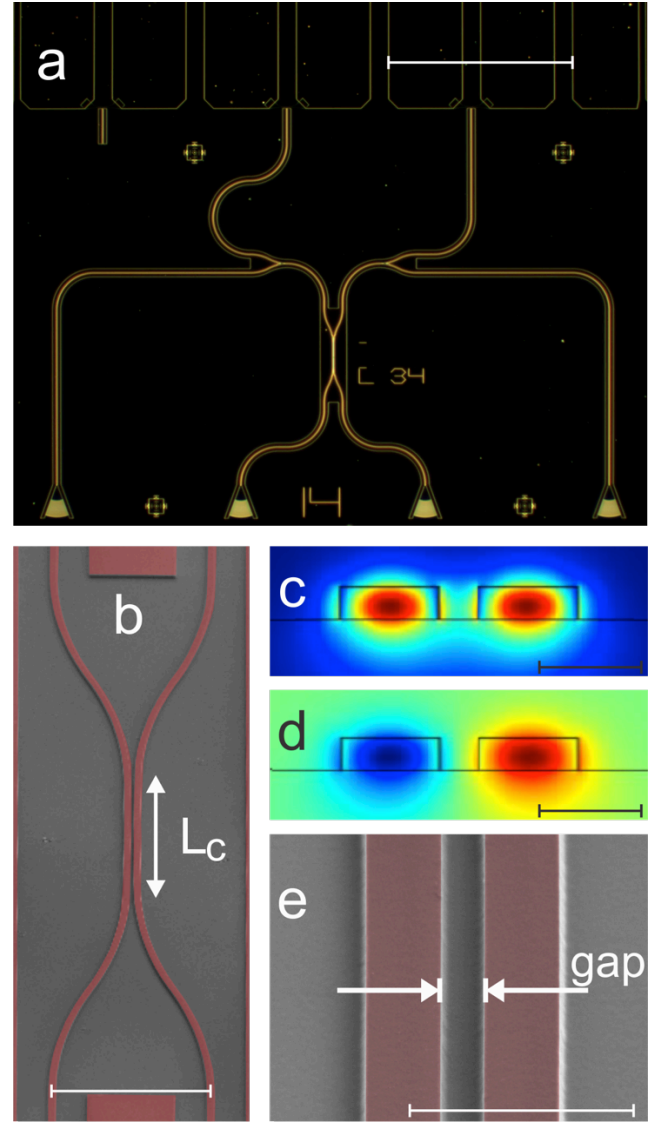

**Fig. 2 Directional coupler device design.** a) dark field micrograph of a HOM-device with two input grating couplers (bottom center), two output grating couplers for device alignment and calibration (bottom left/right), the directional coupler (DC) and waveguide coupled SSPDs, which are contacted via Au-electrode pads (top) (scale bar: 250 μm); b) SEM image of a directional coupler of nominal length $L_c$ made from 1 μm wide SiN waveguides (scale bar: 25 μm); c) FEM simulation of the symmetric TE field mode in the coupling region (max./min.: $1.8 \cdot 10^{10}$ Vm$^{-1}$ / 0 Vm$^{-1}$, scale bar: 1 μm); d) anti-symmetric TE mode (max./min.: $\pm 1.8 \cdot 10^{10}$ Vm$^{-1}$, scale bar: 1 μm); e) center region of the directional coupler where evanescent coupling of the field modes occurs (scale bar: 3 μm).

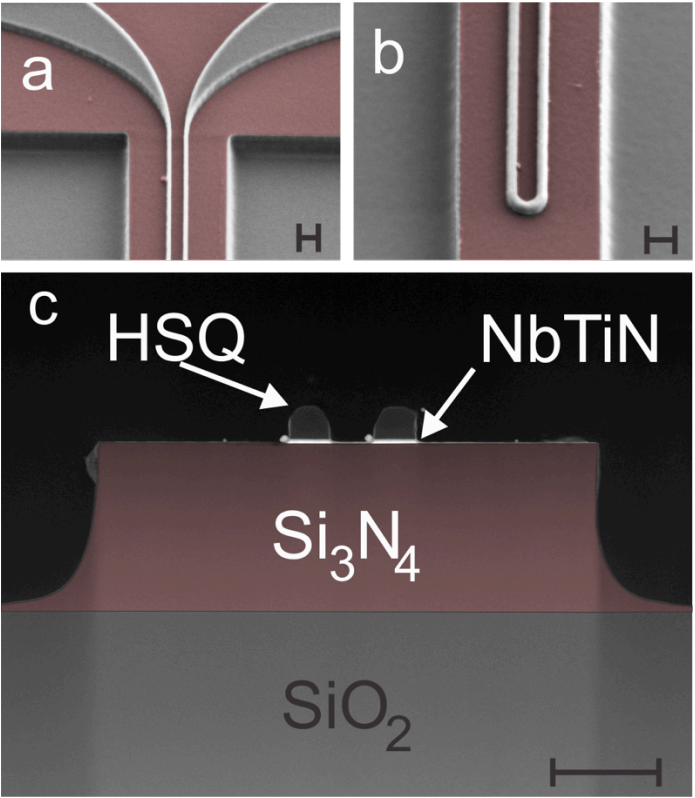

**Fig. 3 Waveguide coupled SSPD.** a) scanning electron micrograph of the leads connecting a 50 nm narrow nanowire SSPD to Au-contact pads (see Fig. 2a); b) U-shaped part of a 40 μm long nanowire SSPD for optimal bias current distribution in the bending region where photons are incident from the SiN-waveguide underneath; c) transmission electron micrograph of the SSPD-waveguide cross section. The 8 nm thin NbTiN nanowire (here 80 nm wide) is covered with electron beam lithography resist (HSQ) and centered on top of a 330 nm high SiN waveguide on a buried oxide layer. All scale bars: 200 nm.

SSPDs were fabricated from sputter deposited 8.2 nm thin NbTiN films (see methods). We pattern 50 nm wide, 40 μm long U-shaped nanowires which are connected via wider leads (Fig. 3a) to the electrode pads (Fig. 2a). We ensure that the SSPDs are precisely centered on top of the waveguides for optimal performance by aligning the nanowires to the same marks, which are subsequently used for patterning the SiN-layer. We confirm that layer alignment between different lithography steps is better than 50 nm in scanning and transmission electron microscopy, as shown in Fig. 3. A calibration of the detector performance yielded an on-chip detection efficiency of 11.5 % and a dark count rate of 0.7 & 2.2 Hz for a typical pair of SSPDs, when biasing close to the critical current (including black body radiation and stray light). We observe that our fabrication process features high yield of functional devices (see supplementary note 2).

**HOM-interference measurement.** To measure quantum interference between photons produced in SPDC we first match the optical delays in the interferometer formed by the fiber PBS and the on-chip waveguide beam splitter (BS, see Fig. 1). We determine the zero delay position using a 2.4 ps pulsed telecom wavelength laser (with corresponding 1.6 mm coherence length) and observing first order interference between split pulses at one of the monitor ports with a fast O/E converter. Using this starting position we then send photon pairs from the SPDC source onto the chip and record detection events for both SSPDs at the outputs of the directional coupler.

For delay positions larger than the coherence length, $\tau_c$, of the down conversion photons we observe a coincidence rate of 4.2 ± 0.1 Hz for 256 ps binning of the photon arrival time data using a nominal 775 nm pump power of 10.5 mW (before coupling to the ppKTP waveguide). As we scan the delay line around the zero delay position we observe how the coincidence rate drops almost to zero from its initial value for unmatched arrival times at the beam splitter and then recovers to its initial value of about 4.2 Hz away from the zero delay position, as shown in Fig. 4a. This is the expected behavior for HOM-interference of temporally correlated indistinguishable photons. We approximate the spectral bandwidth of the down conversion photons incident on the beam splitter by a Gaussian function and fit the raw coincidence rate data with the corresponding function [39]:

$$C = C_n\left(1 - Ve^{-d^2/(2\sigma^2)}\right) \quad (1)$$

where $C_n$ is the coincidence rate for photons with large arrival time delays $\Delta t \gg \tau_c$, $V$ is the HOM-dip visibility, $d$ is the delay position and $\sigma$ the standard deviation. From the fit shown in Fig. 4a we extract the visibility $V = 96.9 \pm 5.3\%$ and a full width at half maximum of $w = 518 \pm 41$ μm, corresponding to a coherence time of $1.7 \pm 0.1$ ps from which we estimate a SPDC photon bandwidth of $2.1 \pm 0.2$ nm [40].

Note, that the HOM-interference visibility was extracted directly from the raw data shown in Fig. 4a. Hence, it contains contributions from accidental coincidence detection events, where each SSPD registered a count within the user-specified coincidence time window although no photon-pair was detected. Such accidental coincidence events occur statistically as photons belonging to different pairs and/or detector dark counts, are registered within a time interval shorter than the coincidence time window. Here we choose a coincidence time window of 256 ps, which is significantly longer than the jitter of our SSPDs (approx. 50 ps [28]) but short enough to avoid a significant background of accidental coincidences: at less than 10 kHz counting rate per SSPD (depending on pump power) we estimate an accidental coincidence contribution of 0.02 Hz, which is contained in the rate shown in Fig. 4a. Our data clearly benefits from the low dark count rates of the two SSPDs used here (0.7 and 2.2 Hz, respectively), which only cause a negligible background contribution to the measured coincidence rate.

We repeat the two-photon interference measurement for a significantly lower pump power of 3.5 mW to avoid higher order processes in the SPDC process but only observe a small improvement of HOM-interference visibility to 97.1 % (see supplementary notes 3). We thus conclude that higher order SPDC processes do not contribute appreciable to the observed coincidence rate at zero delay.

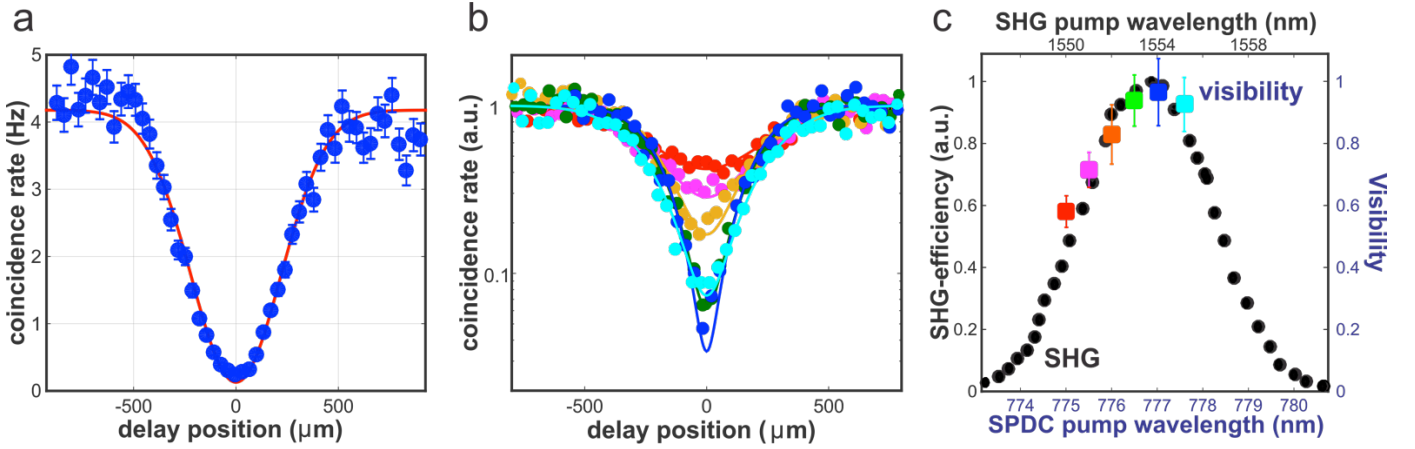

**Fig. 3 On-chip HOM-interference.** a) Coincidence detection rate (raw) between SSPDs coupled to the directional coupler's output waveguides as a function of photon arrival time delay at a nominal pump power of 10.5 mW. Error bars show the standard deviation of the statistical ensemble of coincidence events. As we scan the position of the optical delay line (see Fig. 1) the coincidence rate drops from 4.2 Hz close to zero when photons created in SPDC arrive simultaneously at the on-chip directional coupler. From a Gaussian fit to the data we extract a two-photon interference visibility of $V=0.97\pm0.05$ and the coherence time of the SPDC photons as 1.7 ps; b) measured coincidence rates as a function of relative delay between signal and idler photons for various SPDC-pump laser wavelengths, $\lambda_p= 775 - 777.6$ nm. Error bars are similar to those shown in a) but have been omitted for readability. From a fit to the data we find the visibilities $V_{red}(\lambda_p=775\text{ nm}) = 58\pm5\ \%$, $V_{magenta}(\lambda_p=775.5\text{ nm}) = 71\pm6\ \%$, $V_{orange}(\lambda_p=776\text{ nm}) = 83\pm9\ \%$, $V_{green}(\lambda_p=776.5\text{ nm}) = 94\pm8\ \%$, $V_{blue}(\lambda_p=777.1\text{ nm}) = 97\pm10\ \%$, $V_{cyan}(\lambda_p=777.6\text{ nm}) = 93\pm9\%$; c) HOM-interference visibility as a function of SPDC pump wavelength as extracted from the fits in b), where error bars denote 95% confidence bounds of the fit, and SHG efficiency as a function of pump wavelength at similar temperature of the ppKTP crystal.

The fact that we observe a visibility slightly smaller than 100% at $d = 0$ is thus mainly due to detection events from independent pairs which were created within the coincidence detection window, a slight imbalance in the splitting ratio of our on-chip directional coupler and the statistical photon counting noise. We anticipate that fine-tuning of our fabrication recipes will improve the performance of our photonic circuits and detectors to allow for even higher interference visibilities, which comply with fault-tolerant quantum operations [41].

**Characterization of photon indistinguishability.** To achieve optimal visibility in HOM-interference it is necessary that signal and idler photons are indistinguishable at the beam splitter, not only in arrival time but also in all other degrees of freedom. In nanophotonic implementations, as the one presented here, the spatial overlap between the input modes of the directional coupler is guaranteed by the high lithographic control over the waveguide dimensions on-chip. For the waveguide cross section chosen here (330 nm × 1 µm) only a single TE mode is supported such that other polarization and spatial modes are efficiently suppressed. We thus investigate the remaining spectral distinguishability between the down conversion photons in dependence of SPDC pump detuning, which can be varied off-chip.

The frequency correlation between signal and idler output modes in SPDC is described by a joint spectral amplitude which is given by the product of a pump spectral amplitude and a phase matching function. The latter can be approximated by a Gaussian function with width $\sigma_s$ around the degenerate (quasi-) phase matching frequency, $\omega_0$, where $\Delta\mathbf{k}=0$ with $\omega_s = \omega_i$. If the 775 nm pump frequency, $\omega_p$, is detuned from $\omega_0$, the signal and idler spectra around $\omega_s$ and $\omega_i$, respectively, will shift according to the phase matching conditions imposed by energy and momentum conservation: $\omega_p = \omega_s + \omega_i$ and $\Delta\mathbf{k} = \mathbf{k}_p - \mathbf{k}_s - \mathbf{k}_i - 2\pi/\Lambda$, with poling period $\Lambda$ and wave vectors $k_m = n(\lambda_m, T) \cdot \frac{\omega_m}{c}$. Here $T$ is the SPDC source temperature and the refractive indices $n(\lambda_m, T)$ for pump, signal and idler, $m=p,s,i,$ are given by the Sellmeier equations. A change in pump wavelength thus introduces spectral distinguishability between signal and idler photons, which reduces the visibility in a HOM-interference experiment.

We scan the delay between signal and idler photons around the zero-delay position for pump laser wavelengths of $\lambda$=775-777.6 nm and observe the variation of interference visibility with pump wavelength, shown in Fig. 4b. For better comparison all data is normalized to the coincidence rate at $\Delta\tau \to \infty$ as determined from the respective fit. At $\lambda$=775 nm pump wavelength the phase matching conditions cause the signal and idler spectra to shift significantly such that the interference visibility drops to 58% (see red data in Fig. 4b). As the pump wavelength is increased to 777.1 nm the HOM-interference visibility gradually increases to 97% (see dark blue data in Fig. 4b) before it starts dropping again for $\lambda > 777.1$ nm (see cyan data in Fig. 4b). The variation of interference visibility follows roughly a Gaussian distribution (see Fig. 4c) as expected from shifting the approximately Gaussian signal and idler spectra with respect to each other. High visibility is achieved over a relatively broad spectral range ($V$>90% over $\Delta\lambda \approx 1.5$ nm around $\lambda = 777.1$ nm), which shows that signal and idler photon distinguishability is under accurate control in our experiment.

For comparison we show the second harmonic generation (SHG) efficiency for pump wavelengths $\lambda$=1546-1562 nm of the ppKTP crystal at similar temperature in Fig. 4c. Both SHG efficiency and HOM-interference visibility show similar behavior as a function of the respective pump laser wavelength. This relation between SHG and SPDC phase matching is expected in a crystal of given material properties, waveguide length and cross-section [42].

## Discussion

Recent experiments in quantum optics [3,4,5,43] manifest an ever more pressing need for a scalable solution to integrate photonic circuits and single-photon detectors. In particular, single-photon detection and high-visibility quantum interference have been identified as the two essential requirements for realizing scalable linear optic quantum computation [15]. Here we have shown how SSPDs embedded with nanophotonic circuits address these needs and achieve the requirements for scalable quantum technology on silicon chips. We observe high-visibility quantum interference of photons produced in SPDC with waveguide coupled SSPDs and demonstrate that photon-distinguishability is under accurate control in this architecture. All of the fabrication techniques used here can in principle be adapted to scalable technology developed for the complementary metal oxide semiconductor industry, even at the front-end of a CMOS-line [44]. We anticipate that fine-tuning of superconducting film and photonic circuit parameters and the implementation of photon-number resolving architectures [45,46] will further increase the functionality of our integrated quantum photonic system.

Recent progress in realizing sources of non-classical light directly on silicon chips ideally complements the integration of single-photon detectors and photonic circuits described here. Such integrated quantum light sources were realized employing spontaneous four wave mixing [31,47] and the excitation of waveguide coupled quantum emitters [48,49] but could also be realized via spontaneous parametric down conversion in III-nitride waveguides [50,51]. Combining nanophotonic sources, circuits and single-photon detectors on a silicon chip will allow for generating, processing and detecting quantum information all on one scalable platform.

## Methods

**Device fabrication.** We deposit NbTiN on commercial stoichiometric 330 nm $Si_3N_4$ on 3 μm thermally grown $SiO_2$ on Si wafers. The film-thickness of the NbTiN layer is controlled via timed reactive ion sputtering from an NbTi alloy target in an Ar-$N_2$ atmosphere at room temperature. From atomic force microscopy, transmission electron microscopy (TEM) and square resistance measurements we infer a film thickness of 8.2 nm and a deposition rate of 1.33 nm/s. Transmission electron micrographs (Fig. 3c) on a reference device show that our NbTiN films are slightly thicker than those used in previous device generations [25,26,28]. This could explain the somewhat lower detection efficiency and dark count rate as compared to those reported in ref. 28, which relied on NbTiN-films sputtered elsewhere. We anticipate that fine-tuning of our NbTiN sputter-recipe and film thickness will yield SSPD-performance on a par with previous demonstrations.

After NbTiN-film-deposition we define electrode pads and alignment marks (see Fig. 2a) for subsequent layers in electron beam lithography using double-layer polymethyl methacrylate (PMMA) positive-tone resist. After development in methyl isobutyl ketone and isopropyl alcohol we deposit an 8 nm Ti adhesion layer and 150 nm gold (Au) in electron-beam evaporation followed by lift-off in acetone. In a second high-resolution (100 kV) electron beam lithography step the detector nanowires are patterned in negative-tone hydrogen silsesquioxane (HSQ) resist. Each detector pair is aligned separately to the Au alignment marks in the write-field of the respective device. After development in tetramethylammonium hydroxide based developer the pattern is transferred to the NbTiN layer in a timed reactive ion-etching (RIE) step employing tetrafluoromethane (CF4) chemistry. In a third and final electron beam lithography step we expose the waveguide layer in positive-tone ZEP520A polymer resist. The patterns for each photonic circuit device are aligned to the same alignment marks used in the previous step for defining the respective nanowire detector pair. Following development in xylenes the waveguide patterns are transferred to the SiN film via carefully timed RIE in fluoroform (CHF3). The resulting devices are shown in Fig. 2 and 3.

## Acknowledgements

C.S. acknowledges financial support from the Deutsche Forschungsgemeinschaft (SCHU 2871/2-1). X.S.M. is supported by a Marie Curie International Outgoing Fellowship within the 7[th] European Community Framework Programme. H.X.T acknowledges support from a Packard Fellowship in Science and Engineering and a CAREER award from the National Science Foundation. We thank Dr. Michael Rooks and Michael Power for their assistance in device fabrication.

## Authors contribution

H.X.T. and C.S. conceived the experiment, C.S. and X.G. performed the measurements and analyzed the data, L.R.F. deposited the NbTiN films, C.S. wrote the manuscript, fabricated and characterized the devices, C.S., X.G., M.P., X.S.M., and H.X.T. discussed the results, H.X.T. supervised the work.

## Competing financial interests

The authors declare no competing financial interests.

## Corresponding author

Correspondence to: H.X. Tang.

## References

1. T. D. Ladd, F. Jelezko, R. Laflamme, Y. Nakamura, C. Monroe, J. L. O'Brien, "Quantum Computers", Nature **464**, 45-53 (2010).
2. A. Aspuru-Guzik, P. Walther, "Photonic Quantum Simulators", Nat. Phys. **8**, 285-291 (2012).
3. J. B. Spring, et al., "Boson sampling on a photonic chip", Science **339**, 798-801 (2013).
4. M. A. Broome, et al. "Photonic boson sampling in a tunable circuit ", Science **339**, 794-798 (2013).

5. N. Spagnolo, et al. "Experimental validation of photonic boson sampling", Nat. Photon. **8**, 615-620 (2014).
6. A. Politi, M. J. Cryan, J. G. Rarity, S. Yu, J. L. O'Brien, "Silica-on-silicon waveguide quantum circuits", Science, **320**, 646-649 (2008).
7. J. L. O'Brien, A. Furusawa, J. Vuckovic, "Photonic quantum technology", Nat. Photon. **3**, 687-695 (2009).
8. P. J. Shadbolt, M. R. Verde, A. Peruzzo, A. Politi, A. Laing, M. Lobino, J. C. F. Matthews, M. G. Thompson, J. L. O'Brien, "Generating, manipulating and measuring entanglement and mixture with a reconfigurable photonic circuit", Nat. Photon. **6**, 45-49 (2012).
9. A. Politi, J. C. F. Matthews, J. L. O'Brien, "Shor's quantum factoring algorithm on a photonic chip", Science **325**, 1221 (2009).
10. H. Takesue, S. W. Nam, Q. Zhang, R. H. Hadfield, T. Honjo, K. Tamaki, Y. Yamamoto, "Quantum key distribution over a 40-dB channel loss using superconducting single-photon detectors", Nat. Photon. **1**, 343-348 (2007).
11. D. Grassani, S. Azzini, M. Liscidini, M. Galli, M. J. Stern, M. Sorel, J. E. Sipe, D. Bajoni, "Micrometer-scale integrated silicon source of time-energy entangled photons", Optica **2**, 88-94 (2015).
12. G. E. Moore, "Cramming More Components onto integrated circuits", Electronics **38**, 114-117 (1965).
13. W. Bogaerts, R. Baets, P. Dumon, V. Wiaux, S. Beckx, D. Taillaert, B. Luyssaert, J. van Campenhout, P. Bienstman, D. van Thourhout, "Nanophotonic waveguides in silicon-on-insulator fabricated with CMOS technology", J. Lightwave Technol. **23**, 401-412 (2005).
14. G. N. Gol'tsman, O. Okunev, G. Chulkova, A. Lipatov, A. Semenov, K. Smirnov, B. Voronov, A. Dzardanov, C. Williams, R. Sobolewski, "Picosecond superconducting single-photon optical detector", Appl. Phys. Lett. **79**, 705-707 (2001).
15. E. Knill, R. Laflamme, G. J. Milburn, "A scheme form efficient quantum computation with linear optics", Nature **409**, 46-52 (2001).
16. R. Raussendorf, H. J. Briegel, "A one-way quantum computer", Phys. Rev. Lett. 86, 5188-5191 (2001).
17. F. Marsili, V. B. Verma, J. A. Stern, S. Harrington, A. E. Lita, T. Gerrits, I. Vayshenker, B. Baek, M. D. Shaw, R. P. Mirin, S. W. Nam, "Detecting single infrared photons with 93% system efficiency", Nat. Photon. **7**, 210-214 (2013)
18. R. H. Hadfield, "Single-photon detectors for optical quantum information applications", Nat. Photon. **3**, 696-705 (2009).
19. E. Saglamyurek, J. Jin, V. B. Verma, M. D. Shaw, F. Marsili, S. W. Nam, D. Oblak, W. Tittel, "Quantum storage of entangled telecom-wavelength photons in an erbium-doped optical fiber", Nat. Photon. **9**, 83-87 (2015).
20. F. Bussieres, C. Clausen, A. Tiranov, B. Korzh, V. B. Verma, S. W. Nam, F. Marsili, A. Ferrier, P. Goldner, H. Herrmann, C. Silberhorn, W. Sohler, M. Afzelius, N. Gisin, "Quantum teleportation from a telecom-wavelength photon to a solid-stat quantum memory", Nat. Photon. **8**, 775-778 (2014).
21. X. Hu, C. W. Holzwarth, D. Masciarelli, E. A. Dauler, K. K. Berggren, "Efficiently coupling light to superconducting nanowire single-photon detectors", IEEE Trans. Appl. Supercond. **19**, 336-340 (2009).
22. Sprengers, J. P. et al. "Waveguide superconducting single photon detectors for integrated quantum photonic circuits", Appl. Phys. Lett. **99**, 181110 (2011).
23. C. Schuck, W. H. P. Pernice, O. Minaeva, M. Li, G. Gol'tsman, A. V. Sergienko, H. X. Tang, "Matrix of integrated superconducting single-photon detectors with high timing resolution", IEEE Trans. Appl. Supercond. **23**, 2201007 (2013).
24. W. H. P. Pernice, C. Schuck, O. Minaeva, M. Li, G. Gol'tsman, A. V. Sergienko, H. X. Tang, "High-efficiency, ultrafast single-photon detectors integrated with nanophotonic circuits", Nat. Comm. **3**, 1325 (2012).
25. C. Schuck, W. H. P. Pernice, H. X. Tang, "NbTiN superconducting nanowire detectors for visible and telecom wavelengths single-photon counting on Si3N4 photonic circuits", Appl. Phys. Lett. **102**, 051101 (2013).
26. C. Schuck, W. H. P. Pernice, X. Ma, H. X. Tang, "Optical time domain reflectometry with low noise waveguide-coupled superconducting single-photon detectors", Appl. Phys. Lett. **102**, 191104 (2013).
27. F. Najafi, J. Mower, N. C. Harris, F. Bellei, A. Dane, C. Lee, X. Hu, P. Kharel, F. Marsili, S. Assefa, K. K. Berggren, D. Englund, "On-chip detection of non-classical light by scalable integration of single-photon detectors", Nat. Comm. **6**, 5873 (2015).
28. C. Schuck, W. H. P. Pernice, H. X. Tang, "Waveguide integrated low noise NbTiN nanowire single-photon detectors with milli-Hz dark count rate", Sci. Rep. **3**, 1893 (2013).
29. C. K. Hong, Z. Y. Ou, L. Mandel, "Measurement of subpicosecond time intervals between two photons by interference", Phys. Rev. Lett. **59**, 2044-2046 (1987).
30. A. Zeilinger, "General properties of lossless beam splitters in interferometry", Am. J. Phys. **49**, 882-883 (1981).
31. J. W. Silverstone, D. Bonneau, K. Ohira, N. Suzuki, H. Yoshida, N. Iizuka, M. Ezaki, C. M. Natarajan, M. G. Tanner, R. H. Hadfield, V. Zwiller, G. D. Marshall, J. G. Rarity, J. L. O'Brien, M. G. Thompson, "On-chip quantum interference between silicon photon-pair sources", Nat. Photon. **8**, 104-108 (2014).
32. X. Xu, Z, Xie, J. Zheng, J. Liang, T. Zhong, M. Yu, S. Kocaman, G. Q. Lo, D. L. Kwong, D. R. Englund, F. N. C. Wong, C. W. Wong, "Near-infrared Hong-Ou-Mandel interference on a silicon quantum photonic chip", Opt. Express **21**, 5014-5024 (2013).
33. J. W. Silverstone, R. Santagati, D. Bonneau, M. J. Strain, M. Sorel, J. L. O'Brien, M. G. Thompson, "Qubit entanglement on a silicon photonic chip", Preprint at http://arxiv.org/abs/ 1410.8332 (2014).
34. T. Gerrits, F. Marsili, V. B. Verma, L.K. Shalm, M. Shaw, R. P. Mirin, S. W. Nam, "Spectral correlation measurements at the Hong-Ou-Mandel interference dip", Phys. Rev. A **91**, 013830 (2015).

35. J. S. Fakonas, H. Lee, Y. A. Kelaita, H.A. Atwater, "Two-plasmon quantum interference", Nat. Photon. **8**, 317-320 (2014).
36. R. W. Heeres, L. P. Kouwenhoven, V. Zwiller, "Quantum interference in plasmonic circuits", Nature Nanotech. **8**, 719-722 (2013).
37. T. Zhong, F. N. C. Wong, T. D. Roberts, P. Battle, "High performance photon-pair source based on fiber-coupled periodically poled KTiOPO4 waveguide", Opt. Express **17**, 12019-12030 (2009).
38. C. Wang, B. Lichtenwalter, A. Friebel, H. X. Tang, "A closed-cycle 1K refrigeration cryostat", Cryogenics **64**, 5-9 (2014).
39. H. A. Bachor, T. C. Ralph, "A Guide to Experiments in Quantum Optics", Wiley-VCH (2004).
40. M. Born, E. Wolf, "Principles of optics", Cambridge University Press (1999).
41. A. Laing, A. Peruzzo, A. Politi, M. R. Verde, M. Halder, T. C. Ralph, M. Thompson, J. L. O'Brien, "High-fidelity operation of quantum photonic circuits", Appl. Phys. Lett. **97**, 211109 (2010).
42. L. G. Helt, M. Liscidini, J. E. Sipe, "How does it scale? Comparing quantum and classical nonlinear processes in integrated devices", J. Opt. Soc. Am. B **29**, 2199-2212 (2012).
43. A. Crespi, R. Osellame, R. Ramponi, V. Giovannetti, R. Fazio, L. Sansoni, F. De Nicola, F. Sciarrino, P. Mattioli, "Anderson localization of entangled photons in an integrated quantum random walk", Nat. Photon. **7**, 322-328 (2013).
44. S. K. Tolpygo, V. Bolkhovsky, T. J. Weir, L.M. Johnson, M. A. Gouker, W. D. Oliver, "Fabrication Process and Properties of Fully-Planarized Deep-Submicron Nb/Al-AlO$_x$/Nb Josephson-Junctions for VLSI Circuits", IEEE Trans. Appl. Supercond. **25**, 1101312 (2015).
45. A. Divochiy, F. Marsili, D. Bitauld, A. Gaggero, R. Leoni, F. Mattioli, A. Korneev, V. Seleznev, N. Kaurova, O. Minaeva, G. Gol'tsman, K. G. Lagoudakis, M. Benkhaoul, F. Lévy, A. Fiore, "Superconducting nanowire photon-number resolving detector at telecommunication wavelengths", Nat. Photon. **2**, 302-306 (2008).
46. D. Sahin, A. Gaggero, Z. Zhou, S. Jahanmirinejad, Mattioli, R. Leoni, M. Lermer, M. Kamp, S. Höfling, A. Fiore, "Waveguide photon-number-resolving detectors for quantum photonic integrated circuits", Appl. Phys. Lett. **103**, 111116 (2013).
47. S. F. Preble, M. L. Fanto, j. A. Steidle, C. C. Tison, G. A. Howland, Z. Wang, P. M. Alsing, "On-chip quantum interference from a single silicon ring resonator source", Phys. Rev. Applied **4**, 021001 (2015).
48. G. Reithmaier, S. Lichtmannecker, T. Reichert, P. Hasch, K. Müller, M. Bichler, R. Gross, J. J. Finley, "On-chip time resolved detection of quantum dot emission using integrated superconducting single-photon detectors", Sci. Rep. **3**, 1901 (2013).
49. S. L. Mouradian, T. Schröder, C. B. Poitras, L. Li, J. Goldstein, E. H. Chen, M. Walsh, J. Cardenas, M. L. Markham, D. J. Twitchen, M. Lipson, D. Englund, "Scalable Integration of Long-Lived Quantum Memories into a Photonic Circuit", Phys. Rev. X **5**, 031009 (2015).
50. C. Xiong, W. H. P. Pernice, K. K. Ryu, C. Schuck, K. Y. Fong, T. Palacios, H. X. Tang, "Integrated GaN photonic circuits on silicon (100) for second harmonic generation", Opt. Express **19**, 10462-10470 (2011).
51. C. Xiong, W. H. P. Pernice, X. Sun, C. Schuck, K. Y. Fong, H. X. Tang, "Aluminum nitride as a new material for chip-scale optomechanics and nonlinear optics", New J. Phys. **14**, 095014 (2012).

# Supplementary Information

## Supplementary Note 1

**Calibration devices.** For determining the coupling gap and length, $L_c$, which corresponds to 50:50 coupling between the two waveguides of a beam splitter, we fabricate Mach-Zehnder and beam splitter devices as shown in supplementary figure 1. For the beam splitter devices we inject a 1550 nm wavelength laser at one of the inputs and measure transmission through the device at both outputs of the directional coupler. For the Mach-Zehnder interferometer devices we scan the wavelength of the input laser around 1550 nm wavelength and record the resulting interference fringes at the outputs. Maximum extinction is expected for 50:50 splitting ratio. We fabricate devices of different coupling gap and width on a bare SiN on SiO$_2$ on Si chip, i.e. the material which is used as a substrate for fabricating the PIC-SSPD devices described in the main text. We scan the coupling gap and width around the design values determined from FEM simulations, i.e. 28 µm length for 400 nm gap (see main text).

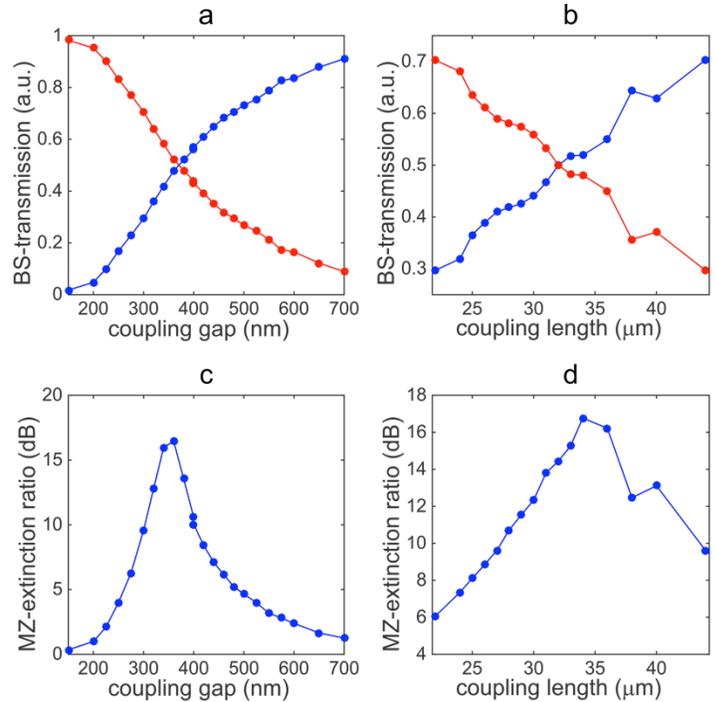

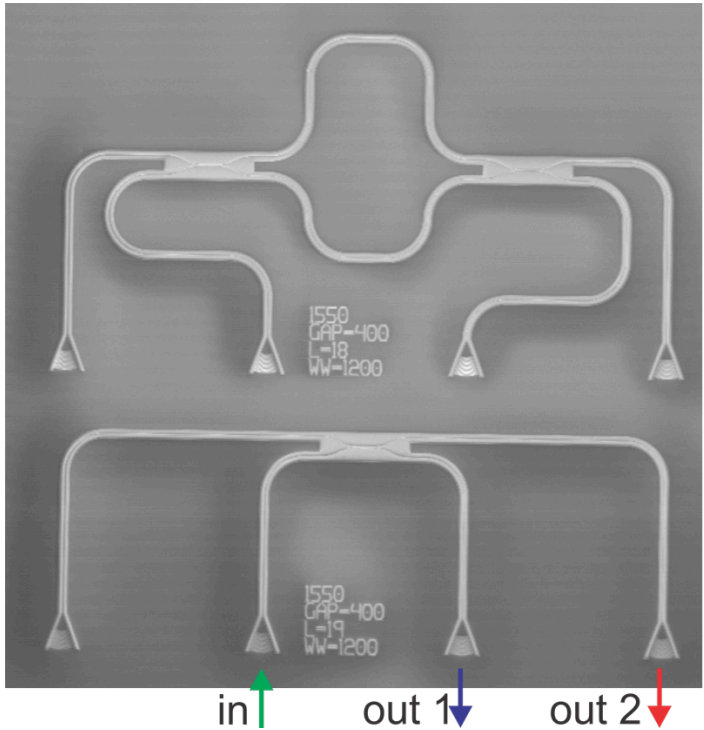

**Supplementary Figure 1: Directional coupler calibration devices.** Scanning electron micrograph of Mach-Zehnder interferometer (top) and beam splitter devices (bottom) to calibrate the directional coupler splitting ratio as a function of coupling length and gap between waveguides of given width and height. Laser light ($\lambda$=1550 nm) is injected at one input port and we detect the transmission at both output ports.

**Supplementary Figure 2: Coupling gap and length calibration.** For the beam splitter (BS) devices shown in supplementary figure 1 (bottom) we record the transmission at the output ports, a) for various coupling gaps 150-700 nm keeping the coupling length constant at 28 µm. 50:50 splitting is observed at 370 nm; b) for various coupling lengths 22-44 µm while keeping the coupling gap constant at 400 nm. 50:50 splitting is observed at 32 µm. For the Mach-Zehnder interferometer (MZ) devices shown in supplementary figure 1 (top) we record the extinction of interference fringes at the output ports, c) for various coupling gaps 150-700 nm keeping the coupling length constant at 28 µm. 50:50 splitting, corresponding to maximal extinction, is observed at 360 nm; d) for various coupling lengths 22-44 µm while keeping the coupling gap constant at 400 nm. 50:50 splitting, corresponding to maximal extinction, is observed at 34 µm.

The results of the calibration measurements are shown in supplementary figure 2. From scanning the coupling gap at (fixed) 28 µm coupling length we find 50:50 splitting for a 370 nm gap between waveguides (supplementary figure 2a). Similarly we observe maximal extinction (limited by the amplified spontaneous emission of the laser used in these measurements), corresponding to 50:50 coupling, in Mach-Zehnder interferometers consisting of two directional couplers both with 360 nm gaps and 28 µm coupling length by design (supplementary figure 2c). When fixing the gap at 400 nm and scanning the coupling length we find 50:50 splitting at 34 µm and 32 µm from the beam splitter transmission (supplementary figure 2b) and Mach-Zehnder interferometer extinction data (supplementary figure 2d), respectively. The mismatch between the design values determined by FEM simulations (see main text) and the measured values is likely due to a different device geometry of the fabricated devices with respect to the simplified design considered in FEM simulations.

Notably, we find very high device yield and similar grating coupler transmission across all devices (up to 6% of the input power through the device), which demonstrate the robustness of our fabrication recipe. These measurements are done in a separate setup at room temperature.

Based on the good agreement of the calibration data from BS and MZ devices we establish a gap of 400 nm and a coupling length of 33±1 μm as design parameters for the directional couplers used in the devices considered in the main text. Slight deviations from 50:50 splitting in PIC-SSPD devices can occur due to the additional processing steps required for SSPD fabrication. E.g. the etching of the NbTiN-layer can slightly reduce the underlying SiN film thickness (see Methods in main text) and thus result in a slight change of waveguide geometry in the coupling region, which in turn causes a slight splitting ratio offset from 50:50. We anticipate that this effect can be accounted for in future device generations.

## Supplementary Note 2

**Device yield.** A typical chip design contains a large number of similar circuits (108 photonic circuits and 324 detectors in our case). For the chips used here we vary a large number of parameters across the devices on a chip, e.g. nanowire width of the SSPDs, grating coupler period, directional coupler gap and length, and other device geometry. To estimate fabrication yield of our circuits and detectors on a chip we perform device characterization at room temperature prior to cryogenic cooling because not all devices could be characterized at low temperature. The photonic waveguide devices are less critical than the nanowire detectors in terms of feature size but occupy a larger area. Optical transmission measurements yield similar results as compared to those presented in supplementary note 1 and indicate high yield (typically above 95%) of the nanophotonic waveguide circuits, as is routinely achieved for PICs (see also ref. [1]). On the other hand, the performance of the SSPDs is of particular interest for evaluating how device yield affects scalability of integrated quantum photonic circuits. We characterized twenty SSPDs at 1.7K and found for all of them similar efficiency compared to that reported in the main text. However, due to the relatively small number of devices studied at low temperature it is difficult to extract meaningful statistics for making statements about how device yield may affect scalability. Instead, we measure the room temperature resistance of all nanowire detectors on a chip and observed the values in supplementary figure 3 a. All detector-nanowires have the same length (40um) but the width is varied from 25 to 90nm. As expected from the design values we find the narrower nanowires to have higher room temperature resistance (red bars) and the wider nanowires with lower room temperature resistance (blue bars). For nanowires with width larger than 30 nm high yield is apparent from supplementary figure 3. For the example of the 198 nanowires with 50nm width (yellow-green bars) we identified 9 devices in supplementary figure 3 a, which clearly deviate from the median room temperature resistance at this width. At least three of these devices were scratched during chip handling in between fabrication steps, which suggest a yield of approximately 97%. To investigate how yield varies between fabrication runs we fabricated a second chip (with similar design but slightly different NbTiN-film thickness) and repeat the room temperature resistance screening. We find the data shown in supplementary figure 3 b, which even shows a slightly higher yield of 98.5% for 50 nm nanowire SSPDs.

However, we want to emphasize that, in general, similar room temperature resistance values are no guarantee for similar detector performance at low temperature but merely an indication. Future studies of fabrication yield of (large numbers of) superconducting nanowires, including careful characterization at cryogenic temperatures, are necessary to establish its influence on scaling integrated quantum photonic. However, our findings show no indications that the scalability of superconducting-photonic circuits would be limited by fabrication yield. In view of the possibility to take advantage of well controlled, automated processes commonly found in the semiconductor industry for future device generations we thus expect that functional large-scale integrated quantum photonic circuits are feasible.

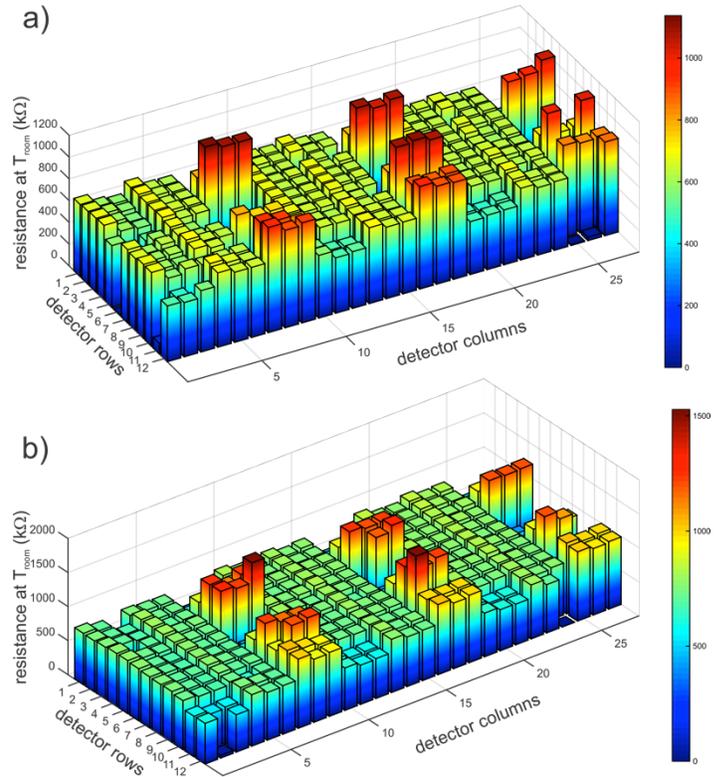

**Supplementary Figure 3: Room temperature resistance of nanowire SSPDs.** a) Room temperature resistance in kΩ (z-axis) for SSPDs on the chip used for the measurements presented in the main text (8.2 nm NbTiN film thickness), ordered in rows & columns on x-and y-axis. Nanowire width varies from 90 nm (e.g. row 1, column 7) to 25 nm (e.g. row 4, column 7); b) room temperature resistance measurement for SSPDs on a chip of similar device layouts but fabricated from 6.7 nm thin NbTiN film.

## Supplementary Note 3

**Power dependence of HOM-interference**. We measure HOM interference for different powers of the 775 nm pump laser to investigate the influence of higher order SPDC-processes on interference visibility. At 10.5 mW pump power we measure a coincidence rate of 4.2 Hz for photons arriving at the on-chip directional coupler with delay $\Delta\tau \to \infty$, as shown in Fig. 4a of the main text. The HOM-interference visibility was determined as 96.9% from a Gaussian fit to the data. We then reduce the 775 nm pump power for SPDC photon pair generation to 3.5 mW and repeat the two-photon interference measurement. In this case we determine a visibility of 97.1% from a Gaussian fit to the data. For direct comparison we show the data of each measurement in supplementary figure 4 normalized to the rate of uncorrelated coincidences (i.e. at $\Delta\tau \to \infty$) determined from the respective fit to the data.

The probability for generating $n$ indistinguishable photon pairs in SPDC is given by [2]:

$$P_n = (1+n)\frac{\left(\frac{\mu}{2}\right)^n}{\left(1+\frac{\mu}{2}\right)^{n+2}} \quad (1)$$

where $\mu = 2\sinh^2(\chi t)$ is the average number of photon pairs produced during a time interval $t$ for a given $\chi$, which is proportional to the second order nonlinear susceptibility and the amplitude of the pump field [3]. The increase in the number of multi-photon-pair processes with pump power is thus almost linear for low μ and monotonically increases with pump power. In turn the visibility of two-photon interference is directly determined by the average number of generated photon pairs and reduces correspondingly with pump power.

$\Delta\tau \to \infty$ The marginal difference of 0.2% in interference visibility observed here for a three-fold increase in pump power shows that the average number of generated photon pairs per $t$=256 ps is low enough such that higher order SPDC-processes do not contribute significantly to the measured coincidence rate (assuming bucket detectors, which do not resolve photon number) in this pump power regime.

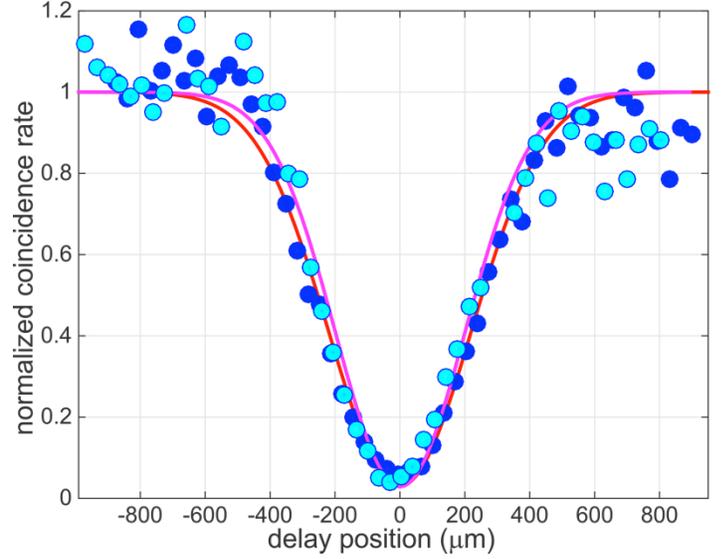

**Supplementary Figure 4: HOM-interference measurement at different SPDC pump power.** Cyan dots: coincidences measured at 3.5 mW power of the 775 nm SPDC-pump laser; magenta curve: fit to 3.5 mW measurement data. Visibility and width of the HOM-dip are found at 97.1% and 472 ± 63 μm, respectively; blue dots: coincidences measured at 10.5 mW power of the 775 nm SPDC-pump laser, for error bars see Fig. 4 in main text; red curve: fit to 10.5 mW measurement. The visibility and width of the HOM-dip are found at 96.9% and 518 ± 41 μm, respectively. Data have been normalized to the respective coincidence rates at $\Delta\tau \to \infty$, as determined from the fit to the 10.5 mW pump and the 3.5 mW pump data.

## Supplementary References

1. W. H. P. Pernice, C. Schuck, O. Minaeva, M. Li, G. Gol'tsman, A. V. Sergienko, H. X. Tang, "High-efficiency, ultrafast single-photon detectors integrated with nanophotonic circuits", Nat. Comm. **3**, 1325 (2012).
2. H. Takesue, K. Shimizu, "Effects of multiple pairs on visibility measurements of entangled photons generated by spontaneous parametric processes," Opt. Commun. 283, 276–287 (2010).
3. D.F. Walls, G. J. Milburn, "Quantum Optics", Springer, Berlin (2008).